\documentclass[nologo,11pt,a4paper]{ETHpaper}
\pdfoutput=1
\usepackage{graphicx, amsmath, amssymb,color,wasysym}
\usepackage{bm}
\usepackage[square,numbers,sort&compress]{natbib}
\usepackage{booktabs}

\begin{document}

\title{The Impact of the COVID-19 Pandemic on Scientific Research in the Life Sciences}

\titlealternative{The Impact of the COVID-19 Pandemic on Scientific Research}

\author{Massimo Riccaboni${}^{1,*}$, Luca Verginer${}^2$}

\address{
${}^1$Piazza S. Francesco, 19, Lucca, 55100, Italy \\
${}^2$Chair of Systems Design, ETH Zurich, Weinbergstrasse 58, 8092 Zurich, Switzerland\\
${}^*$ corresponding author m.riccaboni@imtlucca.it
}

\www{\url{http://www.sg.ethz.ch}}

\makeframing
\maketitle

\begin{abstract}
The COVID-19 outbreak has posed an unprecedented challenge to humanity and science. On the one side, public and private incentives have been put in place to promptly allocate resources toward research areas strictly related to the COVID-19 emergency. But on the flip side, research in many fields not directly related to the pandemic has lagged behind. In this paper, we assess the impact of COVID-19 on world scientific production in the life sciences. We investigate how the usage of medical subject headings (MeSH) has changed following the outbreak. We estimate through a difference-in-differences approach the impact of COVID-19 on scientific production through PubMed. We find that COVID-related research topics have risen to prominence, displaced clinical publications, diverted funds away from research areas not directly related to COVID-19 and that the number of publications on clinical trials in unrelated fields has contracted. Our results call for urgent targeted policy interventions to reactivate biomedical research in areas that have been neglected by the COVID-19 emergency.
\end{abstract}

\date{\today}

\section*{Introduction}

The COVID-19 pandemic has mobilized the world scientific community in 2020, especially in the life sciences~\cite{Fraser2020preprinting,ruiz2020has}.
In the first three months after the pandemic, the number of scientific papers about COVID was fivefold the number of articles on H1N1 swine influenza~\cite{di2020characteristics}.
Similarly, the number of clinical trials for COVID prophylaxis and treatments skyrocketed~\cite{bryan2020innovation}.
Thanks to the rapid mobilization of the world scientific community, COVID-19 vaccines have been developed in record time.
Despite this undeniable success, there is a rising concern about the negative consequences of COVID-19 on clinical trial research with many projects being postponed~\cite{callaway2020covid,de2020increased,padala2020participant}.
According to Evaluate Pharma, clinical trials were one of the pandemic's first casualties, with a record number of 160 studies suspended for reasons related to COVID in April 2020 \cite{evaluatepreview}.
As a consequence, clinical researchers have been impaired by reduced access to healthcare research infrastructures.
Particularly, the COVID-19 outbreak took a tall on women and early-career scientists \cite{inno2020covid,viglione2020women,seitz2020pandemic, gabster2020challenges}.
On a different ground, Shan and colleagues found that non-COVID-related articles decreased as COVID-related articles increased in top clinical research journals \cite{shan2020publication}. Fraser and coworker found that COVID preprints received more attention and citations than non-COVID preprints \cite{Fraser2020preprinting}.
More recently, Hook and Porter have found some early evidence of ``covidisation'' of academic research, with research grants and output diverted to COVID research in 2020 \cite{hookporter2020}. How much should scientists switch their efforts toward SARS-CoV-2 prevention, treatment, or mitigation? There is a growing consensus that the current level of ``covidisation'' of research can be wasteful \cite{bryan2020innovation,abi2020covid,callaway2020covid}.

\begin{figure}
    \centering
    \includegraphics[width=1\textwidth]{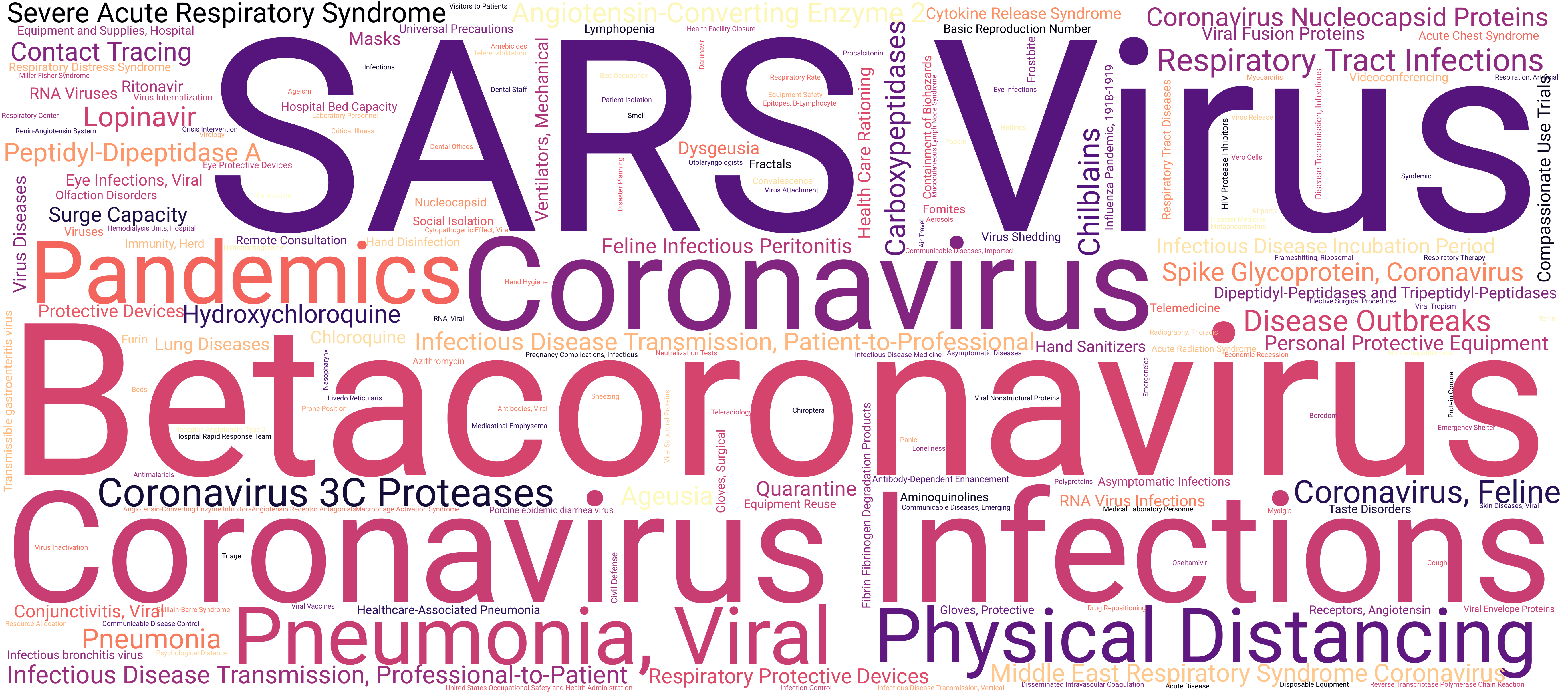}
    \caption{The 200 MeSH terms used most frequently with COVID-specific MeSH Terms are shown as word-cloud. The size of the term is proportional to its relatedness to COVID-19 MeSH terms.}
    \label{fig:wordcloud}
\end{figure}

Against this background, in this paper we investigate if the COVID-19 pandemic has induced a bias in  in biomedical publications toward COVID-related scientific production.
We classify scientific articles of the last three years in PubMed in about 19\,000 ``research fields'' using the curated Medical Subject Headings (MeSH) terminology.
For each research field, we compute a measure of relatedness to COVID-19 (including COVID-19 Testing, Serological Testing, Nucleic Acid Testing, and Vaccines) and SARS-CoV-2 MeSH terms, which first appeared in the scientific literature in December 2019.
A sample of the MeSH terms most closely related to COVID-19 is shown Figure~\ref{fig:wordcloud}.

We first look at how the COVID emergency has caused a profound change in the importance of research topics (as proxied through MeSH) shifting attention to COVID-19 related research.
We then look through a natural-experiment approach on the impact of the pandemic on scientific output.
We consider COVID-19 as a transparent exogenous source of variation across research fields.
By applying a difference-in-differences regression analysis, we find that COVID-related MeSH terms have been much more likely to be published in high-impact journals in the aftermath of the COVID-19 pandemic. The publication bias is even more pronounced for open access research. We also document a significant contraction of non-COVID related clinical trial publications and a negative funding effect for COVID unrelated research fields, with no sign of re-balancing at the end of 2020.

The overall picture that emerges is that there has been a profound realignment of priorities and research efforts within the scientific community and that this shift has displaced unrelated research.

The rest of the paper is structured as follows.
We introduce first the data, our measure of relatedness as well as the econometric specification we rely on to identify the impact of the pandemic on scientific output.
We show the difference in difference analysis highlighting the sudden shift in publications, grants and trial towards COVID related topics and show through a network approach how the MeSH usage patterns of COVID related terms have risen to prominence and also affected the focus of many other fields of research.

\section*{Materials and Methods}

\subsection*{Data}
For the analysis, we use PubMed, specifically the daily updated files up to 31/12/2020.
We consider 3\,360\,248 papers published between January 2018 and December 2020 for the main analysis, as well as 6\,389\,974 papers published from 2015 on-wards to classify grants.
We use SCImago to weigh the papers by the impact factor of the journal they appear in.

\subsection*{Medical Subject Headings (MeSH)}
We rely on the MeSH terminology in this work to approximate ``fields'' of research.
This terminology is a curated medical vocabulary, which is manually added to Papers in the PubMed corpus.
The fact that MeSH terms are manually added makes this terminology ideal for classification purposes.
However, there is a delay between publication and annotation (on the order of months).
To address this delay and have the most recent classification, i.e., December 2020, we search for all 28\,425 MeSH terms using the ESearch facility of PubMed and classify paper by the results.
We apply this method to the whole period (January 2018 to December 2020).
Similarly, we classify papers on clinical trials through ESearch, searching for the specific clinical trial MeSH terms.

\subsection*{COVID-Relatedness}

We estimate the relatedness of a focal MeSH term to COVID-19 MeSH terms as the conditional probability that a paper contains a COVID-19 term given that it includes the focal term. Formally
\begin{equation}
    \text{sim}_i = P( \text{COVID}  | \text{MeSH}_i ) =  P(\text{COVID} \cap \text{MeSH}_i) /  P(\text{MeSH}_i)
\end{equation}

We define a dichotomous variable $\text{COVID}_{i}$ that takes value 1 for MeSH terms highly related to COVID-19 (larger than the threshold value $\tau$) and 0 otherwise.
In our primary analysis we selected the value of $\tau$, which minimizes the pre-2020 distance between the two groups.

\subsection*{Model}
To estimate the impact of the COVID pandemic (treatment) on scientific production, we estimate a random effects panel regression where the units of analysis are about 19\,000 research fields observed over time before and after the COVID-19 pandemic.
Specifically, we estimate the following model
\begin{equation*}
    \ln Y_{it} = \beta_1 + \beta_2 \text{COVID}_{i} + \beta_3 T_{t} + \beta_4 \text{COVID}_{i} \times T_{t}  +
    \nu_i + \epsilon_{it}
\end{equation*}
Where $Y_{it}$ identifies one of the four outcome measures.
The variable $\text{COVID}_{i}$ is 1 if term $i$ is closely related to COVID-19 and 0 otherwise, specifically $\text{COVID}_{i} = 1(\text{sim}_i > \tau)$.
The variable $T_{t}$ identifies the period, i.e., month or year.
$\nu_i$ is the MeSH specific error term, and $\epsilon_{it}$ is the overall error term.
The saturation in $\text{COVID}$ and $T$ means that $\beta_4$ can be interpreted as the Diff-in Diff effect.
In the yearly specification, 2018 is the base year. In the monthly specification, January 2019 is the base month.
The errors are clustered at the MeSH term level.
We drop MeSH terms which have fewer than five papers per month or 60 (=5*12) per year, to exclude infrequent terms.

\subsection*{Additional Robustness Checks}

We have carried out the regression analysis on the unweighted analogs of the dependant variables (i.e., Number of Papers, not IFWN) and raw counts, i.e., not log-transformed.
Moreover, we have carried out the analysis using the continuous COVID-Relatedness measure and a wide range of other $\tau$ values. Finally, we considered an alternative model by dropping MeSH terms with an intermediate level of relatedness to COVID-19.
The results are qualitatively identical.
With the provided code and data all the mentioned robustness checks can be replicated.

\section*{Results}

\subsection*{Impact on Publications, Funding and Clinical Trials}

In January 2020, researchers discovered the cause of a severe respiratory disease of unknown origin: the novel coronavirus SARS-CoV-2. Since then the pandemic has shuffled the research priorities of the worldwide scientific community. We consider SARS-CoV-2 and the related COVID-19 disease as a natural experiment that suddenly affected the scientific community at the beginning of 2020. PubMed scientific publications of the last three years have been classified into about 19\,000 areas of research. We defined a measure of relatedness to COVID-19 as the probability that a paper in a given domain contains SARS-CoV-2 or COVID-19 MeSH terms. Research fields have been divided into two groups: COVID-related areas of research (treated) and other research fields (control). Different thresholds of relatedness to COVID-19 have been used to separate the two groups.

\begin{table}
    \centering
    \caption{Panel Regression}\label{tab:regression}
    {
\def\sym#1{\ifmmode^{#1}\else\(^{#1}\)\fi}
\begin{tabular}{l*{4}{c}}
\toprule
Impact weighted No.\ of
                  &\multicolumn{1}{c}{(1)}&\multicolumn{1}{c}{(2)}&\multicolumn{1}{c}{(3)}&\multicolumn{1}{c}{(4)}\\
scientific publications:                    &\multicolumn{1}{c}{ln(Total)}&\multicolumn{1}{c}{ln(OA)}&\multicolumn{1}{c}{ln(Grants)}&\multicolumn{1}{c}{ln(Trials)}\\
\midrule
Covid-Related     &    0.001         &    -0.008         &     -0.020        &      0.012         \\
                    &      (0.02)         &     (-0.13)         &     (-0.30)         &      (0.09)         \\
[1em]
2019 (Other)          &       0.222\sym{***}&       0.182\sym{***}&       0.203\sym{***}&       0.297\sym{***}\\
                    &    (106.11)         &     (52.38)         &     (54.61)         &     (52.97)         \\
[1em]
2019 $\times$ Covid-Related&     -0.011         &     -0.020         &      0.020         &     -0.033         \\
                    &     (-1.19)         &     (-1.26)         &      (1.40)         &     (-0.86)         \\
[1em]
2020 (Other)         &       0.194\sym{***}&       0.126\sym{***}&      -0.121\sym{***}&      -0.102\sym{***}\\
                    &     (49.23)         &     (25.91)         &    (-23.38)         &    (-15.19)         \\
[1em]
\textbf{2020 $\times$ Covid-Related}&       0.617\sym{***}&       0.919\sym{***}&       0.622\sym{***}&       1.165\sym{***}\\
                    &     (24.75)         &     (29.96)         &     (20.91)         &     (12.01)         \\
[1em]
Constant            &       7.396\sym{***}&       6.499\sym{***}&       5.621\sym{***}&       4.139\sym{***}\\
                    &    (580.26)         &    (473.59)         &    (364.94)         &    (240.27)         \\
\midrule
Observations        &       52,597         &       52,597         &       52,597         &       52,597         \\
MeSH Terms          &       18,926         &       18,926         &       18,926         &       18,926         \\
$R^2$ within                &       0.204         &       0.137         &       0.140         &       0.127         \\
$R^2$ overall                &     0.002         &     0.003         &     0.004         &     0.004         \\
COVID Relatedness ($\tau$)                &    0.05          &    0.05          &   0.05          &        0.11      \\
\bottomrule
\multicolumn{5}{l}{\footnotesize \textit{t} the statistics in parentheses are clustered at MeSH term level.}\\
\multicolumn{4}{l}{\footnotesize \sym{*} \(p<0.05\), \sym{**} \(p<0.01\), \sym{***} \(p<0.001\)}
\end{tabular}
}

\end{table}

\begin{figure}
    \centering
    \includegraphics[width=.6\textwidth]{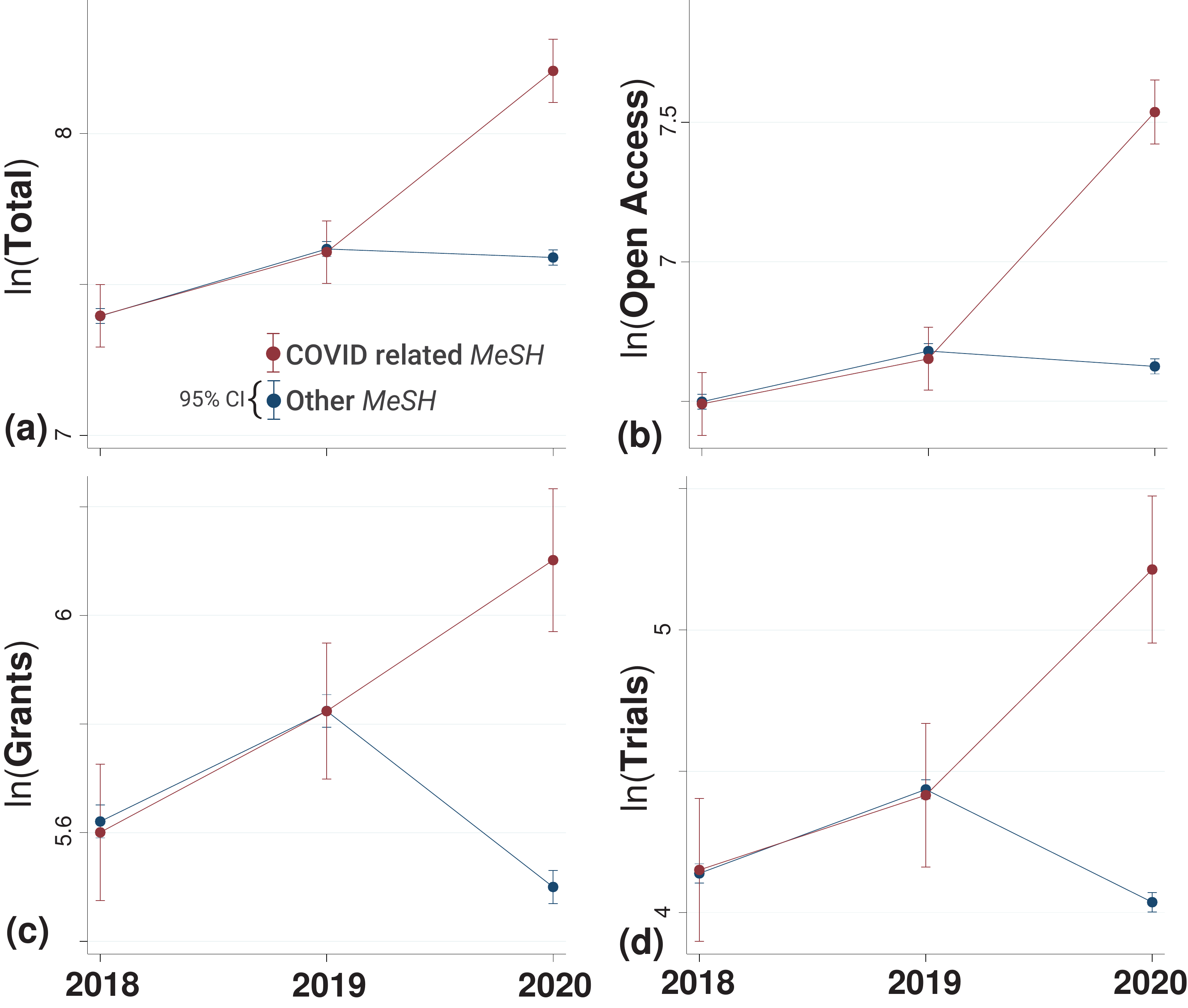}
    \caption{Results of the difference-in-differences regression analysis on the impact of relatedness to COVID-19 and Sars-Cov-2 (treatment) on the scientific outcome in MeSH research domains defined as (1) the yearly impact factor weighted number (IFWN) of publications in years 2019 and 2020 as compared to 2018. A similar analysis is performed for (2) the IFWN of open access publications in PubMed; (3) the IFWN of publications that acknowledged grants and (4) the IFWN of publications related to clinical trials. The figures show that the common trend assumption is satisfied up to 2019. In 2020 there has been a significant increase in COVID-related scientific productions. Non-COVID research lagged behind. All outcome measures are in natural logarithm.
    }\label{fig:yearly_margins}
\end{figure}

In Table~\ref{tab:regression} we report the result of a difference-in-differences regression to estimate the impact of COVID-19 on the journal impact factor weighted number (IFWN) of publications, open access publications, articles related to clinical trials, and publications with grants.
And in Figure~\ref{fig:yearly_margins} we show the margins of this regression, which highlight that the common trend assumption is satisfied since the scientific output grew at the same rate from 2018 to 2019 across both groups.
Then in 2020 the two groups significantly diverged: COVID-related publications experience a significant increase, whereas publications in other fields halted, with a marked drop of clinical trials publications and publications listing grants.

\begin{figure}
    \centering
    \includegraphics[width=.6\textwidth]{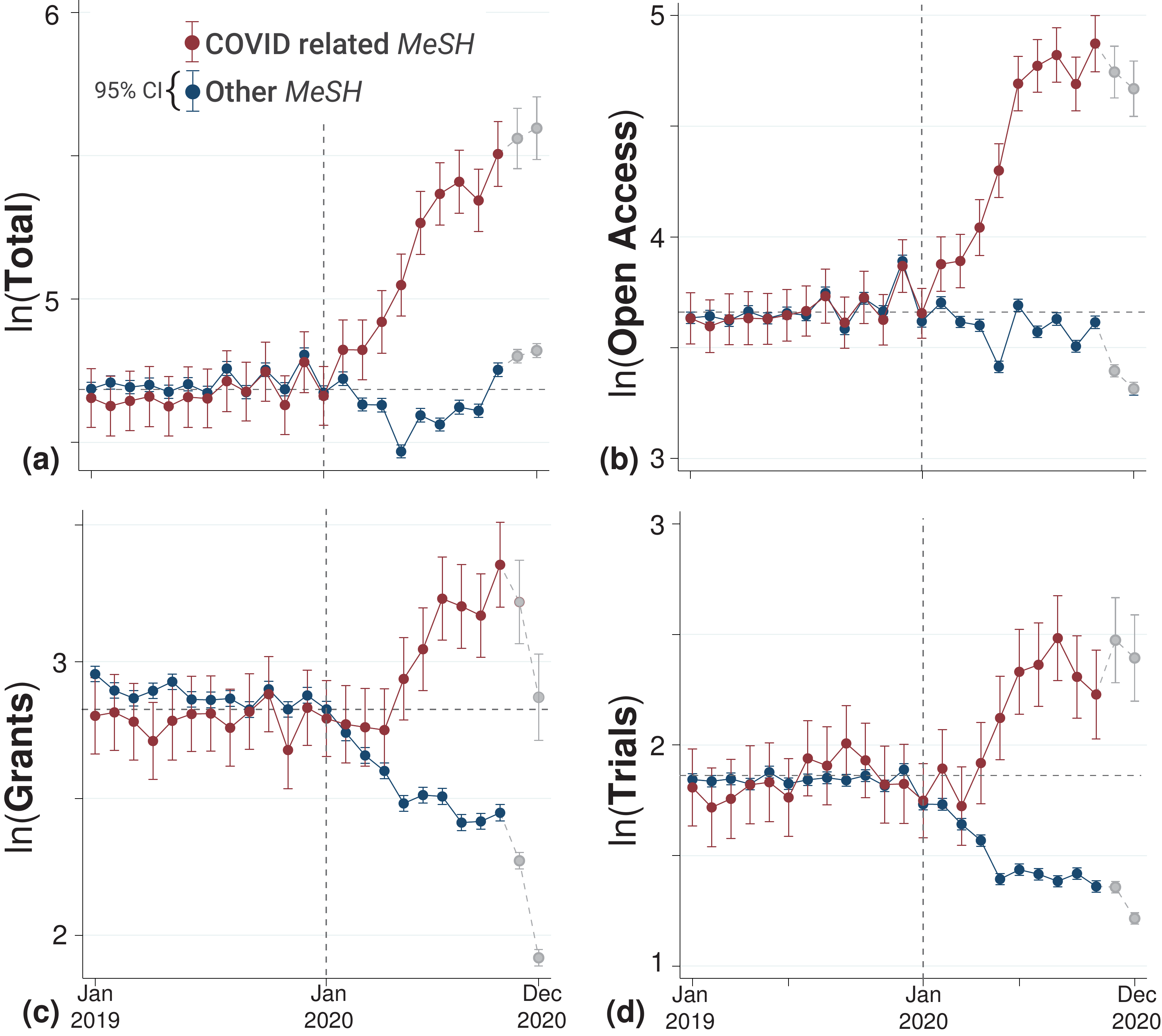}
    \caption{Sub-figures (a-d) are the monthly analogs to the marginals in \ref{fig:yearly_margins}.
    The last two months (Nov and Dec 2020) are grayed out to highlight that MeSH terms and other manually added metadata are added with a delay by PubMed.
    }\label{fig:monthly_margins}
\end{figure}

\begin{figure}
    \centering
    \includegraphics[width=.6\textwidth]{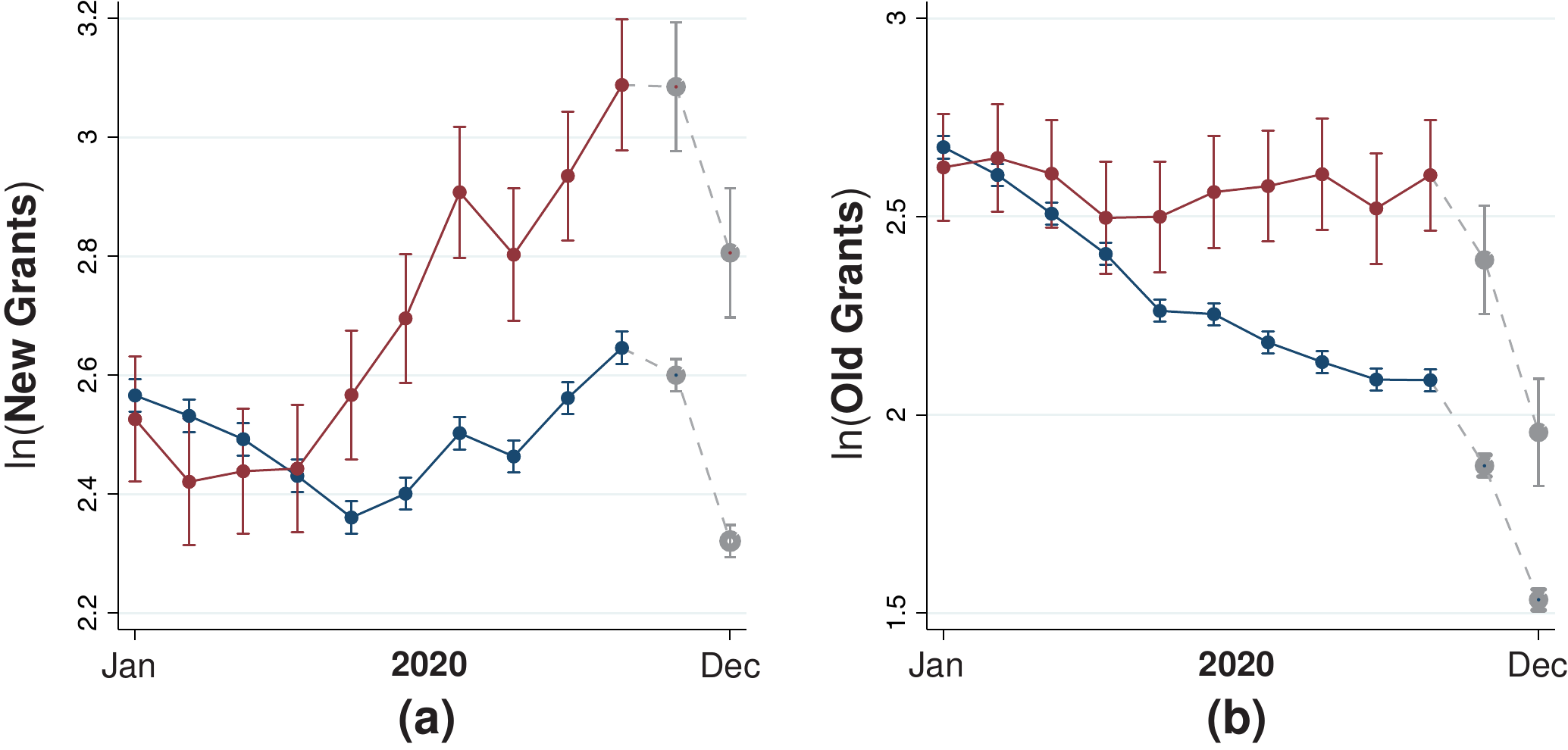}
    \caption{Sub-Figure (a) and (c) show respectively the log of the IWTN of papers listing a \emph{new grant} (first appearance in 2020) and an \emph{old grant} (appeared before 2020)}
    \label{fig:grant_types}
\end{figure}

Figure~\ref{fig:monthly_margins} shows a similar analysis on a monthly basis. The monthly analysis reveals that total scientific output and open access publications in non-COVID research areas bounce back to pre-COVID-19 levels after a dip in May.
Conversely, we do not detect any sign of recovery for publication related to non-COVID clinical trials and granted publications.
To understand if non-COVID grants have been diverted to COVID-related research, we classify the type of grants listed in the publications as ``new grants'' (appear first in 2020) and ``old grants'' (appeared before 2020).
Obviously, old grants could not have been meant for COVID research.
The usage of these two types of grants by the two grups is shown in Figure~\ref{fig:grant_types}.
We notice that old grants are used disproportionately more often in COVID-related areas of research.
This implies that grants which were not meant to sustain COVID research have been diverted to COVID publications.
This is on top of the ad-hoc financial support by new grants for COVID-19 in 2020.

\subsection*{Changes in MeSH usage}

The sudden surge in publications on COVID-19 related terms documented above has also affected the way MeSH terms are used and, by extension, the Biomedical research landscape's focus.
To highlight this shift in focus, we proxy the usage and changing interdependence of MeSH terms by looking at their co-occurrence on all publications in the years 2018, 2019 and 2020.
Precisely, we extract the co-occurrence matrix of all 28,000 MeSH terms as they appear on the 3.3 million papers by year.
These networks summarise their complex interdependence and usage by the scientific community.
We proxy the changes in this network by looking at the change in PageRank centralities across years.
The PageRank centrality tells us how likely a random walker traversing a network would be found at a given node if he follows the edges.
Specifically, for the case of the MeSH co-occurrence network, this number represents how often an author would include a given MeSH term following the observed general usage patterns.
It is a simple measure to capture the complexities of biomedical research. Nevertheless, it captures far reaching interdependence across MeSH terms as the measure uses the whole network to determine the centrality of every MeSH term.  
A sudden change in the rankings and thus the position of MeSH terms in this network suggests that a given research topic has risen from obscurity to prominence as it is used more often with other important MeSH terms (or vice versa).

To show that COVID-related research has profoundly impacted the way MeSH terms are used, we compute for each MeSH how many ranks it gained from 2019 to 2020.
To have a caparison, we compute the gain for the pre-COVID period from 2018 to 2018.

\begin{figure}[!hb]
    \centering
    \includegraphics[width=0.6\textwidth]{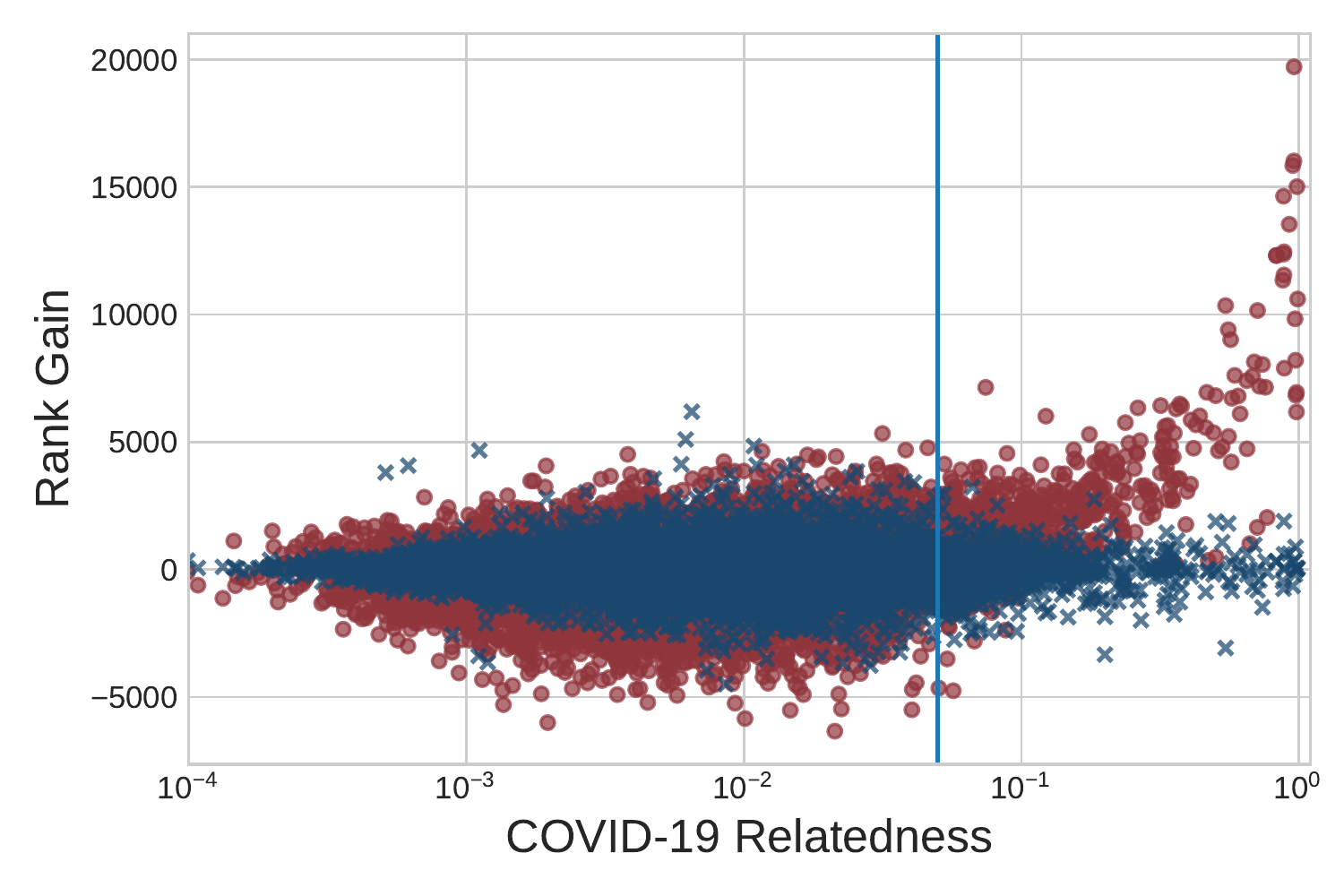}
    \caption{MeSH term importance gain (PageRank) and their COVID relatedness. The red dots show for each MeSH term how many ranks it has gained from 2019 to 2020 as a function of its relatedness to COVID-19 MeSH terms. Similarly, the rank-gain for the period 2018 to 2019 as a reference is shown. The vertical blue line at 0.05 relatedness is the cut-off we use in our preferred specification of the Difference in Difference regression.}
    \label{fig:rank_gain}
\end{figure}

In figure Figure~\ref{fig:rank_gain} we see that MeSH terms with high COVID-19 similarity have risen quickly from obscurity to become \emph{central} to much of the published research.
We find that the Pearson correlation between rank gain and COVID relatedness is very high at 0.5.
We also note that this effect was completely absent in 2019, as indicated by the blue point cloud with a correlation of 0.003.
Note that the same effect can be replicated for other centrality measures (i.e., betweenness)

\section*{Discussion}

The scientific community has swiftly reallocated research efforts to cope with the COVID-19 pandemic, mobilizing knowledge across disciplines to find innovative solutions in record time.
We document this both in the scientific output as well as in the usage of MeSH terms of the scientific community.
The flip side of this sudden and energetic prioritization of effort has caused a sudden contraction of scientific production in other relevant areas of research.

Our results provide compelling evidence that research related to COVID has indeed displaced scientific production in other research fields, with a significant drop of scientific output related to non-COVID clinical trials and a marked reduction of financial support for publications not related to COVID. The displacement effect is persistent to the end of 2020.
Heading into 2021, as vaccination progresses, we highlight the urgent need for science policy to re-balance support for research activity that was put on pause because of the COVID-19 pandemic.

We find that COVID-19 dramatically impacted clinical research.
Reactivation of clinical trials activities that have been postponed or suspended for reasons related to COVID is a priority that should be considered in the national vaccination plans by including patients and researchers in the high priority groups.
Moreover, since grants have been diverted and financial incentives have been targeted to sustain COVID-19 research leading to an excessive entry in COVID-related clinical trials and the ``covidisation'' of research, there is a need to reorient incentives to basic research and otherwise neglected or temporally abandoned areas of research.
Without dedicated support in the recovery plans for neglected research of the COVID-19 era, there is a risk that more medical needs will be unmet in the future, possibly exacerbating the shortage of scientific research for orphan and neglected diseases, which do not belong to COVID-related research areas.

\bibliographystyle{sg-bibstyle-nourl}
\setlength{\itemsep}{0pt}
\small 
\bibliography{references}

\end{document}